\documentclass[10pt,letterpaper]{article}
\usepackage{opex3}
\RequirePackage{amsmath}
\usepackage{graphicx}
\usepackage[breaklinks=true,colorlinks=true,linkcolor=blue,urlcolor=blue,citecolor=blue]{hyperref}
\usepackage[usenames,dvipsnames]{xcolor}
\usepackage{float}
\usepackage[T1]{fontenc}
\usepackage[utf8]{inputenc}
\usepackage{subcaption}
\begin{document}

\title{Detecting atoms trapped in an optical lattice using a tapered optical nanofiber}
\author{T.~Hennessy$^{1,2}$ and Th.~Busch$^{1}$}
\address{$^1$Okinawa Institute of Science and Technology Graduate University, Okinawa, Japan}
\address{$^2$Physics Department, University College Cork, Cork, Ireland}
\email{tara.hennessy@oist.jp}

\begin{abstract}
Optical detection of structures with dimensions smaller than an optical wavelength requires devices that work on scales beyond the diffraction limit. Here we present the possibility of using a tapered optical nanofiber as a detector to resolve individual atoms trapped in an optical lattice in the Mott Insulator phase. We show that the small size of the fiber combined with an enhanced photon collection rate can allow for the attainment of large and reliable measurement signals.
\end{abstract}




\section{Introduction}

Developing tools to control all degrees of freedom of single quantum particles is one of the fundamental aims of the area of quantum engineering. Over the past few decades precision spectroscopy has made significant strides towards achieving this goal for the internal degrees of freedom of atoms and ions, and more recently electro-magnetic trapping technologies have made similar advances in controlling the external degrees. By today; magnetic, optical and magneto-optical traps can  be designed to trap and control large or small numbers of atoms, and even single particles. One example of the latter are optical microtraps, in which single atoms can be localised to an area with dimensions smaller than an optical wavelength. Such traps can, for example, be based on highly focussed laser beams or small scale interference patterns, with the most famous example of the latter being optical lattices. Taking advantage of the existence of the so-called Mott transition at low temperatures, such lattices allow for the creation of periodically spaced arrays of individually trapped atoms in one, two or three dimensions. 

The ability to control the spatial position of single atoms to a high degree of precision is, for example, important in controlling distance dependent interactions. It is therefore necessary to develop diagnostic tools that can measure the position of a single atom with very high fidelity. A number of technologies have been developed during recent years and among these are de-convolution algorithms \cite{Meschede:09} and atom microscopes \cite{Bakr:09}. The latter rely on an advanced optical setup and the ability to position optical elements very close to the atoms. In this work we suggest another method which makes use of technology developed in the emerging area of sub-wavelength diameter optical fibers by calculating the spatial emission profile of an array of periodically spaced two-level atoms in the presence of such a fiber. Optical nanofibers, which are created by placing industrial grade silica fiber over a hot flame and pulling both ends (for a review of this technology see~\cite{ward:2014}), support only a small number of modes and can therefore be thought of as cavities into which enhanced emission rates can be achieved \cite{two atoms,NayakHakuta:08, 6Hakuta:07,Yalla:12, Fujiwara:11,multiatom}. The emission characteristics of an atom into a nanofiber in general depends on the radius of the fiber, the distance from the atom, the wavelength of the transition and the orientation of the atomic dipole \cite{sondergaard:01}. It has recently been shown that a single-mode fiber of radius $200$ nm can collect up to 28\% of the spontaneous emission of a cesium atom when the atom is sitting at the fiber surface \cite{KienHakuta:05}. In larger fibers, which support higher order modes as well, even higher collection rates can be achieved \cite{Masalov:13, nicchor:14-2}. 
 
Here we study the coupling between a row of atoms and the guided modes of a perpendicularly aligned optical nanofiber and show that such a setup can resolve the position of single atoms on length scales that are typical for optical lattices. It can also allow for the detection of  empty sites in a Mott insulator state or be used in a time-dependent way to identify, for example, edge states \cite{topedgestates}. In the following we will first briefly introduce the geometry of the setup (Section~\ref{sec:Potentials}), then present the expressions for the collected radiation (Section~\ref{sec:EmissionRates}) and finally discuss the obtained results (Section~\ref{sec:Results}).

\section{Optical Lattices and nanofibers}
\label{sec:Potentials}

Let us first briefly review the components of the suggested setup. Optical lattices are formed by pairs of counter-propagating lasers which interfere to create spatially periodic arrays of microtraps by employing the dipole force of the standing wave laser light field \cite{latticebook}. The simplest case of an optical lattice trapping potential is given by a one-dimensional model, in which two counter-propagating laser beams interfere. This results in a standing wave for the optical intensity given by $ I(z)=I_0\sin^2(kz)$, where $k=2\pi/\lambda$ is the free space wave number of the laser light and $I_0$ is the maximum intensity of the laser beam. The periodicity of the intensity is $\lambda/2$ and the spatially varying ac Stark shift then forms a potential for the induced dipole moment, ${\bf d}$, of the atom given by 
\begin{equation}
\label{eq:OpticalForce}
U_\text{dip}=-\frac{1}{2}\langle{\bf d\cdot E}\rangle=-\frac{1}{2\epsilon_0c}\operatorname{Re}(\alpha)I.
\end{equation}
Here $\epsilon_0$ is the vacuum permittivity, $c$ is the speed of light and $\alpha(\omega_L)$ is the optical polarizability, which depends on the frequency of the laser field, ${\bf E}$, see \cite{Jackson}. By using light which is blue-detuned ${(\omega_L > \omega_0)}$ or red-detuned ${(\omega_L < \omega_0})$ with respect to the atomic transition $\omega_0$, the atoms can be forced to gather at the nodes or anti-nodes of the laser intensity pattern, respectively. Higher dimensional lattices can simply be created by introducing pairs of counter propagating lasers in the other spatial directions \cite{Greiner:02}, and different spatial geometries can be achieved by varying the angle between these beams \cite{Becker:09}. Optical lattices typically have lattice constants in the range between $400$ nm and $700$ nm and throughout this work we assume a lattice spacing of $\lambda/2=640$ nm.  The atom we consider is $^{133}$Cs, whose ground state has its dominant ($D_2$) line at $\lambda_0$=852 nm.  

Cold atoms trapped in optical lattices provide an adaptable quantum system in which a variety of matter-wave quantum phenomena can be engineered and observed. Notably, it has been possible to induce a quantum phase transition from a superfluid to a Mott insulator state by controlling the tunneling interaction between different sites. This was first demonstrated experimentally in 2002 \cite{Greiner:02}, when a Mott Insulator state with a well-defined number of atoms per lattice site was achieved using $^{87}$Rb. However, until recently it was only possible to demonstrate this phase transition by detecting the loss of coherence using a time-of-flight interferometric measurement, and only the development of so-called atom microscopes has made it possible to resolve single sites inside the lattice \cite{Bakr:09}. The technological difficulty in constructing a device that can resolve single lattice sites lies in overcoming the diffration limit and allowing for sub-wavelength resolution. The atom microscope does this by using advanced optical elements to focus the light to below the diffraction limit.

\begin{figure}[t]
  \centering
  \begin{subfigure}{.4\textwidth}
    \centering
    \includegraphics[width=0.7\linewidth]{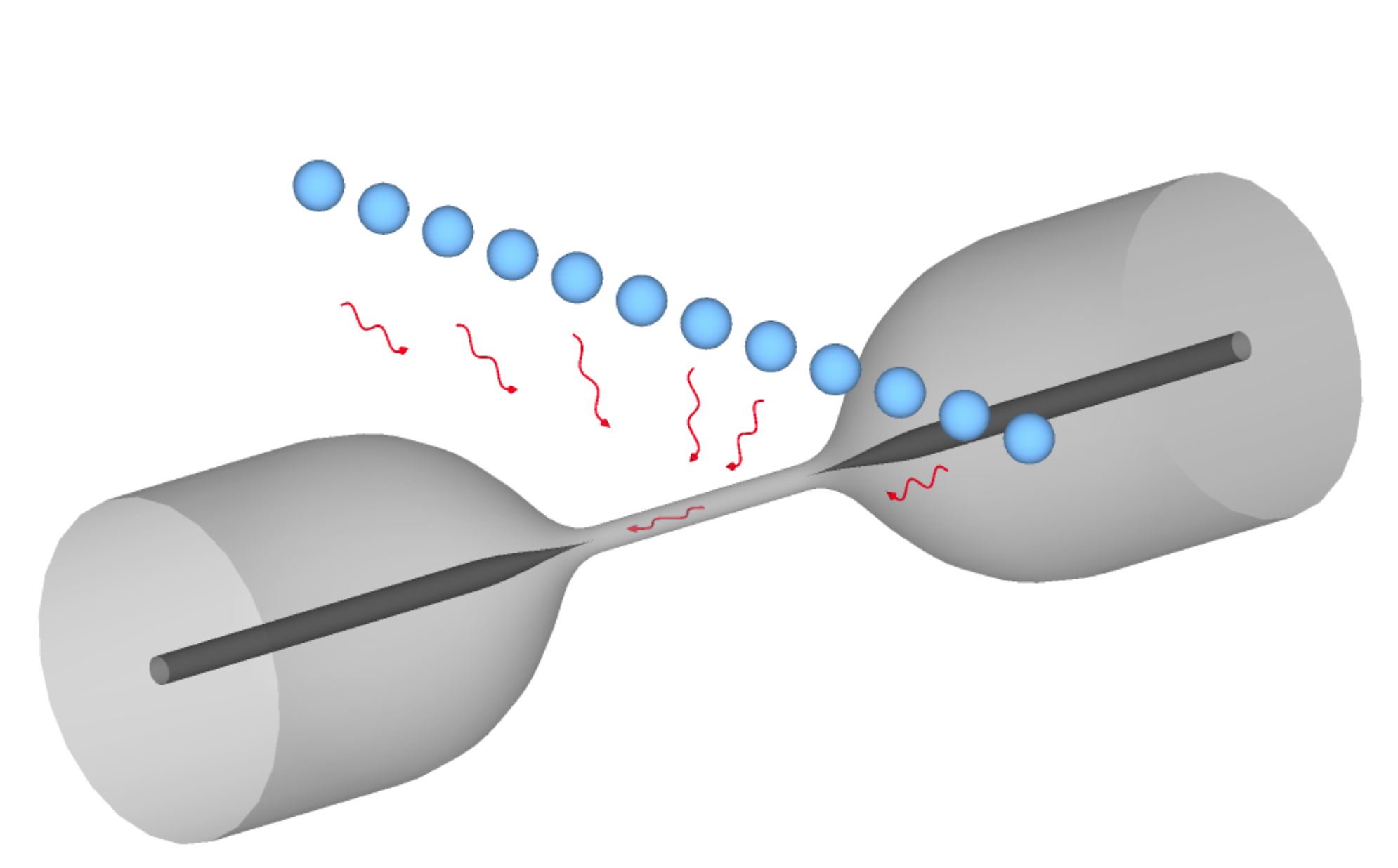} 
    \caption{}
      \label{fig:fiberschematic}
  \end{subfigure}%
  \begin{subfigure}{.6\textwidth}
    \centering
    \includegraphics[width=1\linewidth]{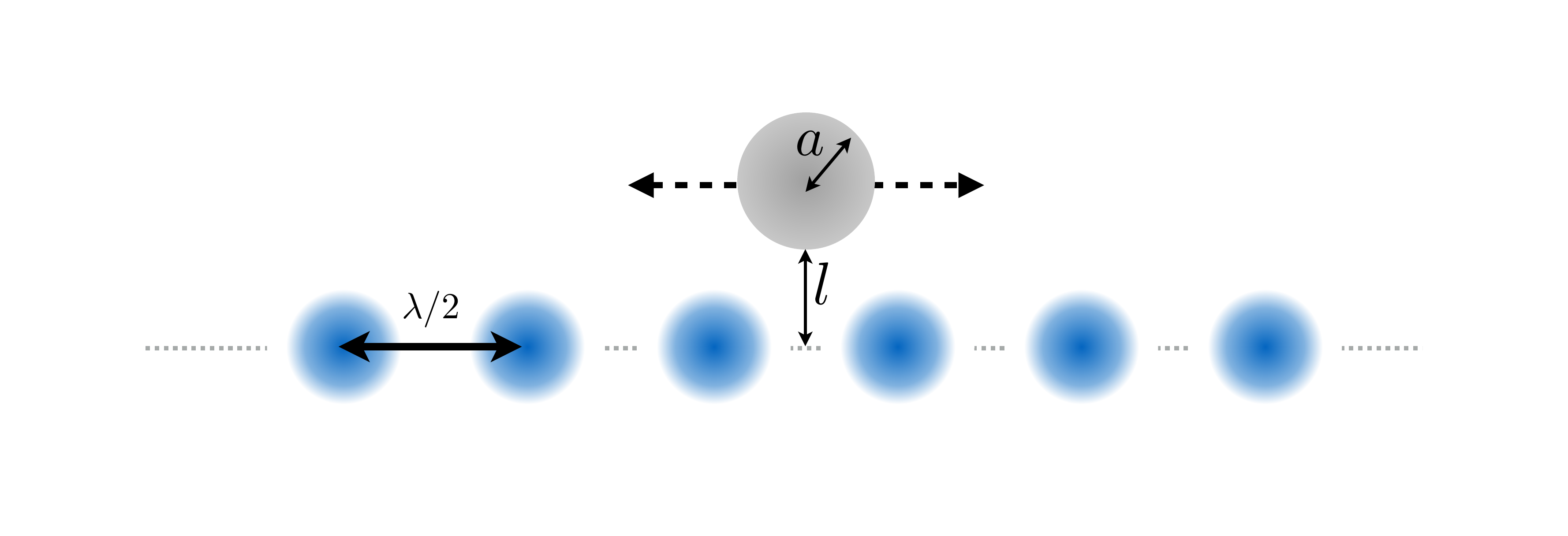} 
        \caption{}
    \label{fig:fiberschematic2}
  \end{subfigure}
 \caption{ (a) Schematic of a tapered optical nanofiber aligned perpendiculary to a periodic array of trapped atoms. In reality the untapered ends have a radius on the order of 125 $\mu m$ and the nanofiber waist is on the order of hundreds of nanometers. The typical length of the tapered region is between 3 and 10 mm, depending on the tapering technique employed~\cite{ward:2014}. (b) Top-view of the fiber and atom configuration. The dipoles are aligned in line and perpendicular to the fiber. Not drawn to scale. }
\label{fig:schematicsidebyside}
\end{figure}

A second strategy for going beyond the diffraction limit is to use sensors of subwavelength size and here we study an approach of this kind based on  collecting the optical emission  using the guided modes of a nanofiber. Recent advances in technology have made it possible to produce tapered fibers with radii as small as a few hundred nanometres \cite{Tong:03, ward:2014}, in which only a very small number of modes can propagate. These devices can be fully integrated in ultracold atom experiments \cite{Vetsch:09, Morrissey:09} and we suggest a geometry in which a fiber is aligned perpendicularly to a row of regularly spaced atoms (see Fig.~\ref{fig:schematicsidebyside}) in order to collect the atomic fluorescence. Due to the small radius of the fiber, such a setup allows for a good spatial resolution of the atomic distribution. \

As already mentioned above, the emission characteristics of an atom into a nanofiber depends on the radius of the fiber, the distance between the atom and the fiber, the wavelength of the transition and the orientation of the atomic dipole~\cite{sondergaard:01}. In the following we will investigate the use of  single- (supporting only HE$_{11}$) and multi-mode fibers (supporting HE$_{11}$, HE$_{21}$,  TE$_{01}$,  TM$_{01}$), as the latter are known to allow for much higher collection rates~\cite{Masalov:13, nicchor:14-2}.  We consider a one-dimensional optical lattice in which  $^{133}$Cs atoms are individuallytrapped with a separation distance $\lambda /2=640$~nm and assume that all dipoles are aligned. For any position of the fiber we therefore only include a contribution from the component of the dipole along the line connecting the fiber centre and the atom. The distance between the surface of the fiber and the axis of the row of atoms is given by $l$ and the fiber radius by $a$ (see Fig.~\ref{fig:fiberschematic2}). The fiber we consider is made of silica, which for a photon of $\lambda_0=852$ nm has a refractive index of $n_1$ = 1.4525 \cite{Sellmeier}.

Following closely the approach of \cite{Masalov:13} to calculate the emission rate into the fiber, let us first consider the emission rate into free space, 
\begin{equation}
  W_{0}=\frac{1}{4 \pi \varepsilon_0} \frac{4 d ^2 \omega_0^3 }{3 \hbar c^3} \;,
\end{equation}
where $d$ is the matrix element of the atomic dipole moment, $\varepsilon_{0}$ is vacuum permittivity and $\omega_{0}$ is the atomic frequency. 
This emission can excite four guided modes given by the four possible combinations of $\pm\sigma$  polarization and  $\pm z$ propagation direction in the fiber. Normalizing the emission rate into each of these modes with respect to the full emission rate, we arrive at 
\begin{equation}
    W_\text{guid}=W_0 \frac{3 {\lambda_{0}} ^2 \beta' }{8 \pi} |{\bf{ E}}|^2\;,
 \label{rateofemission}
\end{equation}
where $\bf{E}$ is made up of the mode functions of the electric parts of the relevant mode outside of the nanofiber. The parameter $\beta'$  is the reciprocal of the group velocity and calculated as $\beta' =d \beta / d k$, where the propagation constant $\beta$
is determined by ensuring continuity between the field components outside of the nanofiber ($r>a$) with the ones at the surface of the fiber~\cite{fiber books}.
%
This leads to an implicit equation that 
has to be solved numerically as a function of the system parameters. 

\begin{table}[t]
  \centering
  \begin{tabular}{c c c}
  \hline
  Fiber Radius&$\beta$&$\beta$ '\\
\hline
  150~nm&7.471$\times 10^6$&1.115\\
  200~nm &7.883$\times 10^6$&1.364\\
  250~nm &8.436$\times 10^6$&1.517\\
  300~nm &8.919$\times 10^6$&1.566\\
  \hline
  \end{tabular}
\caption{Numerical values of $\beta$ and $\beta '$ for the  fundamental HE$_{11}$ mode for the $\lambda_{0}=852$ nm transition in $^{133}$Cs for a fused silica fiber with $n_1$ = 1.4525.}
\label{Tab:PropConst}
\end{table}

\section{Emission rates into four modes of interest}
\label{sec:EmissionRates}

Substituting the electric field components of the relevant modes into Eq.~\eqref{rateofemission} gives four expressions for the rates of emission into the four modes of interest to us \cite{Masalov:13}.  These are the fundamental HE$_{11}$, TE$_{01}$, TM$_{01}$ and HE$_{21}$ modes, each propagating into either $\pm z$ with circular polarisation $\pm\sigma$

\begin{align}
W_{\text{HE}_\text{11}}(r)=&W_0 \frac{3 \lambda^2 \beta' }{8 \pi^2 a^2}\left( \frac{1}{n_1^2 N_1+n_2^2 N_2}\right) \frac{J_1^2 (h a)}{K_1^2 (q a)} \nonumber \\
&\qquad\times \left[ K_1^2 (q r) +\frac{\beta^2}{2 q^2} \left[(1-s)^2 K_0^2 (q r)+(1+s)^2 K_2^2 (q r)  \right]\right],
\label{spontemissionHE11}\\
W_{\text{TE}_\text{01} }(r)=&W_0 \frac{3 \lambda^2 \beta'   }{8 \pi^2 q^2 a^4}\left(\frac{1}{n_1^2 P_1 +n_2^2 P_2 }\right)  K_1^2 (q r) ,
\label{spontemissionTE01}\\
W_{\text{TM}_\text{01}}(r)=&W_0\frac{3\lambda^2\beta'}{8 \pi^2 a^2}\left(\frac{1}{n_1^2 Q_1 +n_2^2 Q_2 }\right)\frac{\beta^2}{q^2} K_0^2 (qr) K_1^2 (qr),
\label{spontemissionTM01}\\
W_{\text{HE}_\text{21}}(r)=&W_0\frac{3\lambda^2\beta'}{8 \pi^2  a^2}\left(\frac{1} {n_1^2 R_1 +n_2^2 R_2 }\right)\frac{J_2^2 (ha)}{K_2^2 (qa)} \nonumber\\ 
&\qquad\times \left[K_2^2(qr)+\frac{\beta^2}{2q^2}\left[(1-u)^2 K_1^2 (qr)+(1+u)^2 K_3^2 (qr)\right] \right]. 
\label{spontemissionHE21}
\end{align}
 In the above equations the $J_n$ are Bessel functions of the first kind, the $K_n$ are modified Bessel functions of the second kind, $a$ is the fiber radius and
\begin{equation}
  q=\sqrt{\beta^2-{n_2}^2 {k}^2}~~~~\text{and}~~~~ h=\sqrt{n_1^2 k^2-\beta^2}.
  \label{qandh}
\end{equation}
The explicit expressions for the constants $N, P, Q$, $R$, $s$ and $u$ are given in the appendix.

The number of modes a nanofiber in vacuum can support is related to the quantity $V=\frac{a \omega_0 }{c} \sqrt{n_1^2(\omega_0)-1}$ \cite{fiber books} and for $V<2.405$ the single-mode condition is satisfied. This means that only the HE$_{11}$ mode can travel in the fiber and for $^{133}$Cs, with $\lambda_0=852$ nm, this corresponds to keeping the fiber radius in the region where $a<309.6$ nm. The range of radii for which the first four modes are supported is  $0.363  \lambda_0 < a < 0.580  \lambda_0$,  which corresponds to  $309.6$ nm $< a < 494.2$ nm.

\section{Results}
\label{sec:Results}
To show that the spatially inhomogeneous emissions rate originating from a row of atoms trapped in a one-dimensional optical lattice can be resolved using a nanofiber, we calculate the emission rate into the fiber as a function of the position of the fiber with respect to the position of the atoms. Since this rate depends on the distance from the atoms, spatial resolution with high visibility can be expected if the fiber is close to the atoms.

\begin{figure}[t]
        \begin{subfigure}{.5\textwidth}
                    \includegraphics[width=\linewidth]{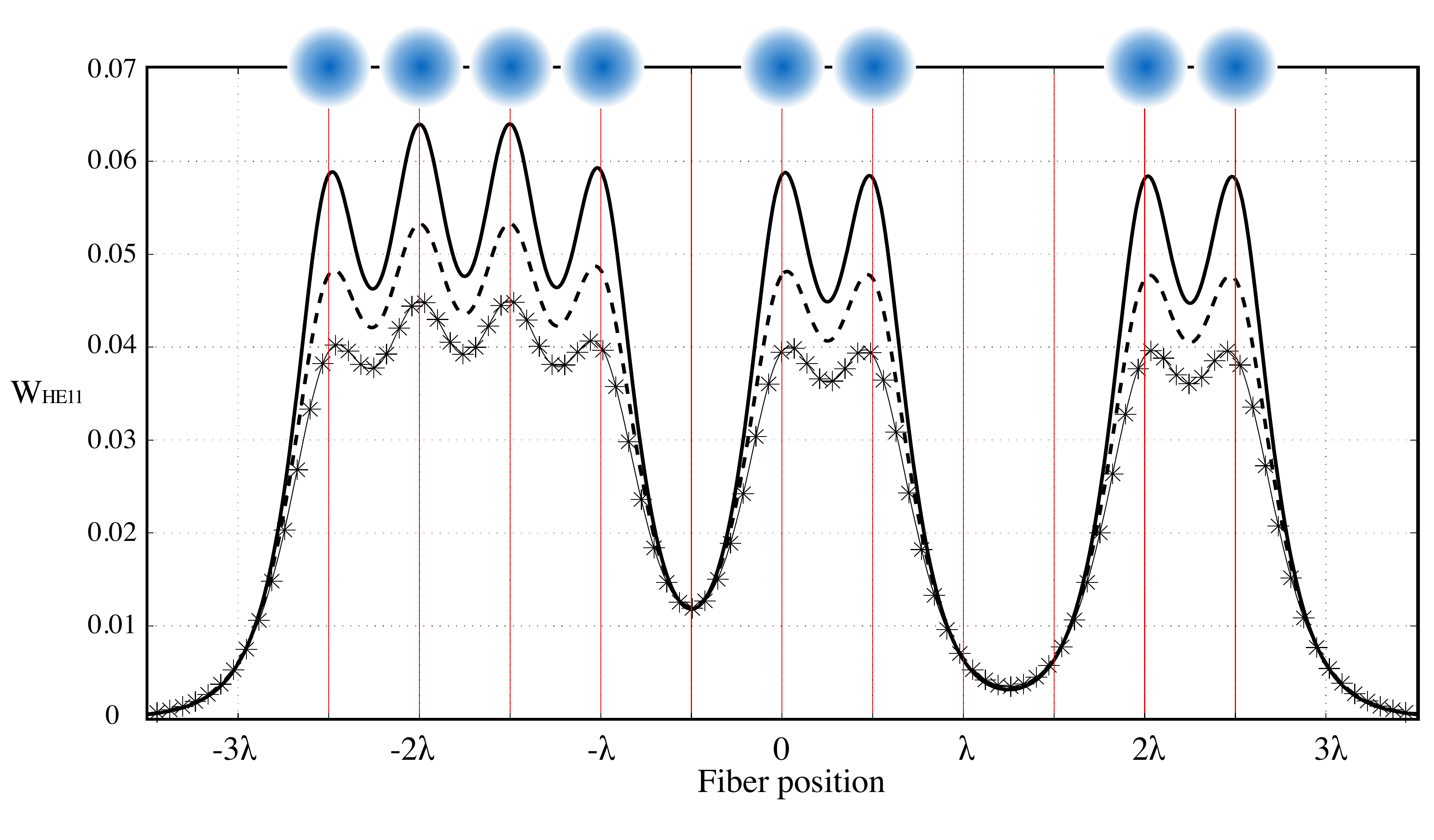} 
                    \caption{Varying the atom-fiber distance}
                    \label{fig:changingd}
        \end{subfigure}
        \begin{subfigure}{.5\textwidth}
            \includegraphics[width=\linewidth]{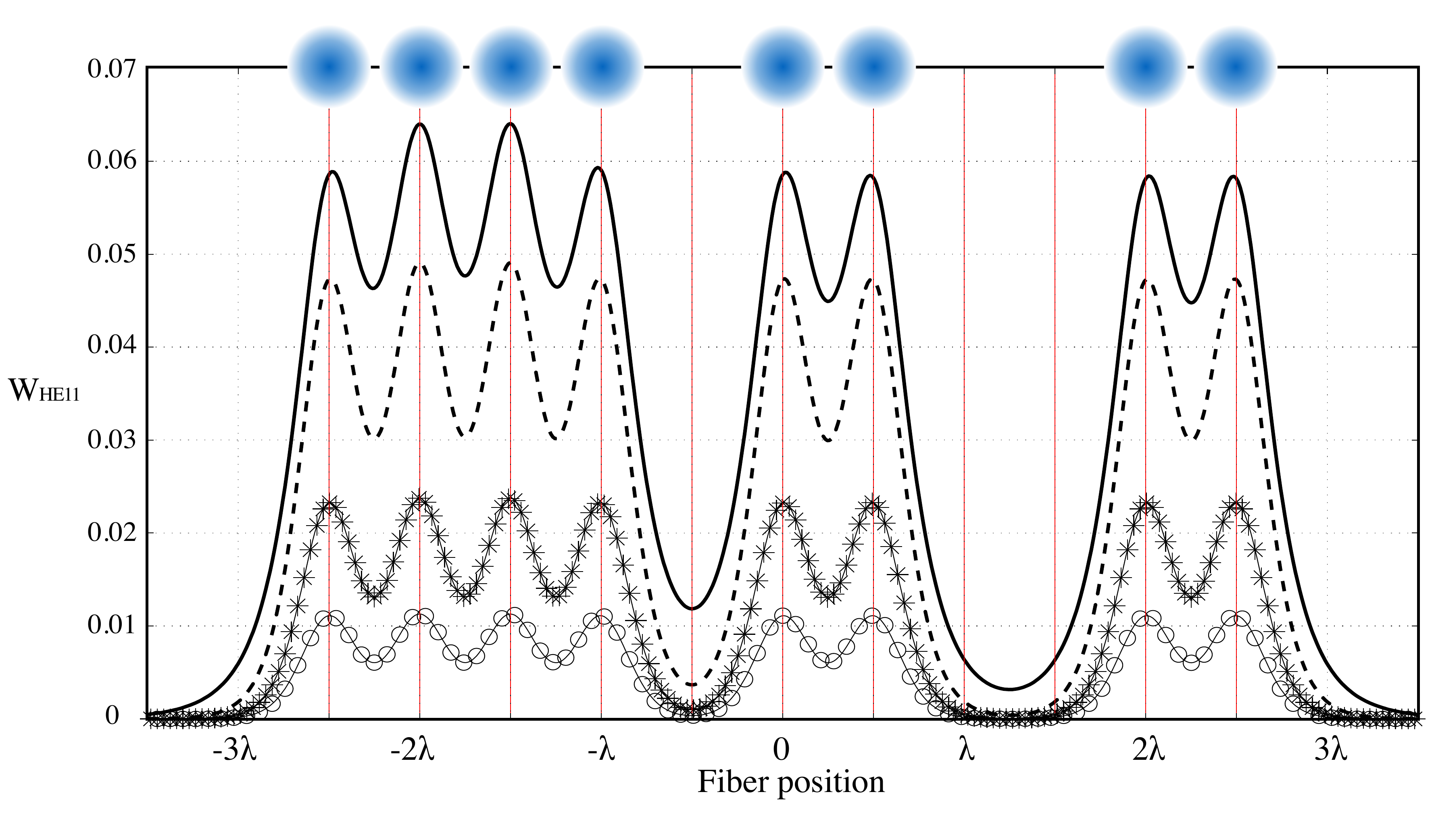} 
            \caption{Varying the fiber radius}
            \label{fig:changinga}
        \end{subfigure}
        \caption{(a) Emission rate into the HE$_{11}$ mode of a fiber with fixed fiber radius $a=150$ nm at a distance, $l$, of 200 nm (solid line), 250 nm (dashed line) and 300 nm (starred line) from the row of atoms. (b) Emission rate into the same mode, but for fibers with different radii at fixed atom-fiber distance $l=200$ nm. The fiber radius is 150 nm (solid line), 200 nm (dashed line), 250 nm (starred line) and 300 nm (circled line).}
        \label{fig:sidebyside}
\end{figure}

We first consider the use of single-mode fibers, which supports the HE$_{11}$ mode only. This means that we are restricted to fibers with a radius of $a<309.6$ nm and in Table \ref{Tab:PropConst} the numerical values for $\beta$ and $\beta '$ are given for fibers of radii 150 nm, 200 nm, 250 nm and 300 nm. The resulting emission rates into a fiber with $a=150$ nm for various distances $l$ between the fiber surface and the row of atoms are shown in Fig.~\ref{fig:changingd} and distinct maxima are clearly visible whenever the fiber is aligned with an occupied trapping position of the lattice (indicated by the vertical red lines and schematic blue atoms). 
 
 As expected, the visibility can be seen to decrease with increasing distance of the fiber from the row of atoms, but even for $l=300$ nm the signal still allows to distinguish individual maxima. Beyond that, when the atoms are a distance away from the fiber surface such that the light from two polarised dipoles  overlaps, it becomes very difficult to resolve the atoms. The positions where no maximum is visible have been intentionally left empty, and the effective extinction of the signal shows that this setup is able to resolve defects in the atomic crystal. As the absence effectively measures a signal homogeneous in space, no degrading of the signal is visible for the parameters shown in the plot.

The dependence of the emission rate into the fiber on the radius of the fiber is shown in Fig.~\ref{fig:changinga}. Note that we keep the distance between the atom row and the fiber surface constant ($l$), which means that as the fiber radius increases, the critical distance from the fiber axis to the atom ($l+a$) increases, and the rate of emission into the guided modes can be expected to reduce.  This is clearly visible in Fig.~\ref{fig:changinga} and the results show that even rather big single-mode nanofibers can record very distinct signals if they are close enough to the atoms. 

\begin{table}[t]
  \centering
  \begin{tabular}{l c c}
  \hline
  Mode&$\beta$&$\beta$ '\\
\hline
  HE11 &9.559$\times 10^6$&1.564\\
  TE01 &8.197$\times 10^6$&1.616\\
  TM01 &7.893$\times 10^6$&1.464\\
  HE21 &7.737$\times 10^6$&1.599\\
\hline
  \end{tabular}
  \caption{Numerical values of $\beta$ and $\beta '$ for a fiber of radius $a=400$~nm fiber for the first four guided modes, again at the $\lambda_{0}=852$~nm transition in $^{133}$Cs.}
\label{Tab:PropConst2}
\end{table}

\begin{figure}[tb]
 \centering
\includegraphics[width=4 in]{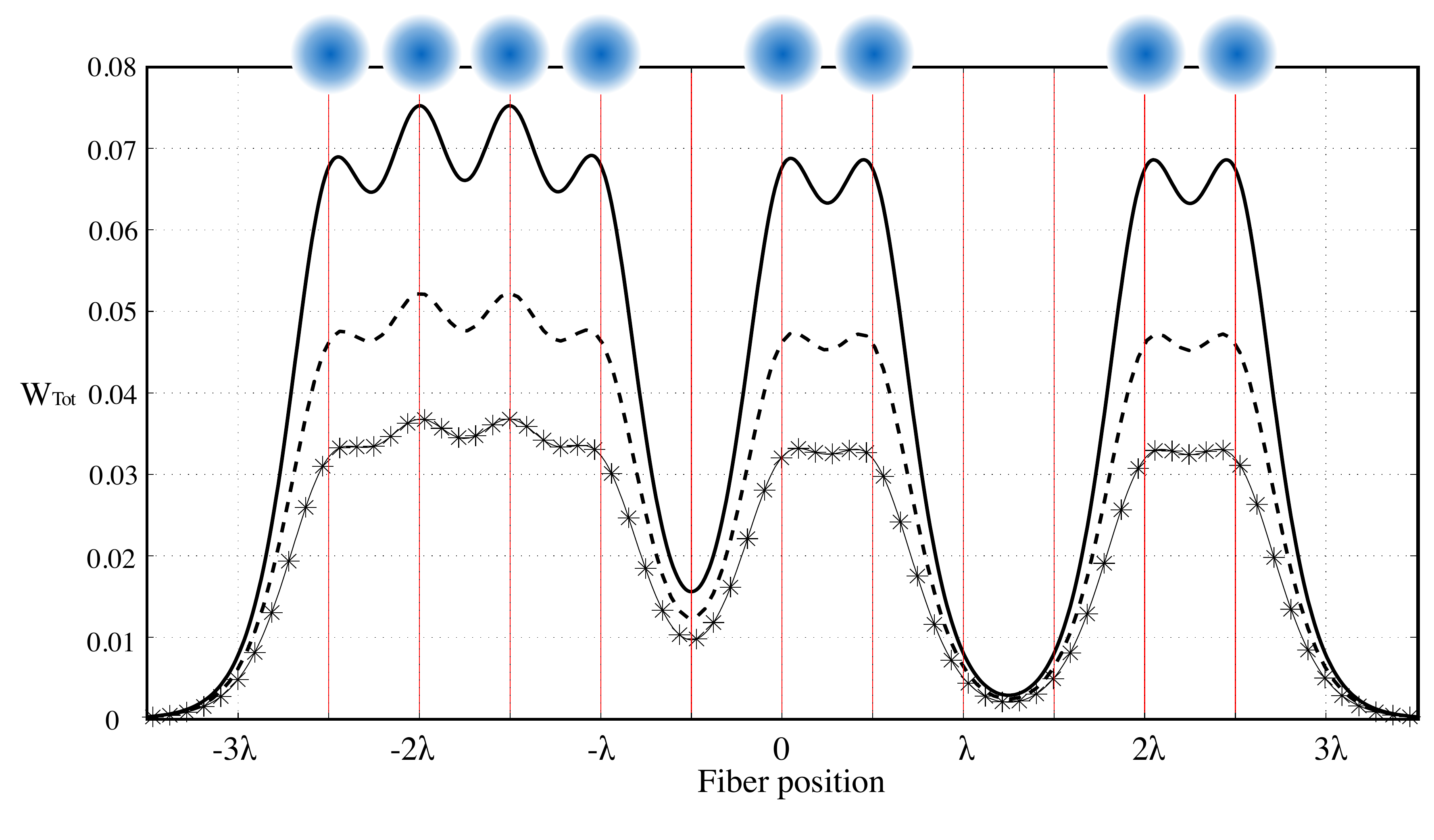} 
\caption{Combined emission rates into the four available modes in a fiber of radius 400~nm at a distance of 200~nm (solid), 250~nm (dashed) and 300~nm (starred). The overall rates of emission are higher than for a single-mode fiber, however, the visibility decreases faster with increasing distance between the atoms and the fiber.}  
\label{fig:fourmodes}
\end{figure}

In slightly larger fibers the presence of three extra modes allows for even higher collection rates and in Fig.~\ref{fig:fourmodes} we show the emission rates into a fiber of radius $a=400$~nm. The relevant values for $\beta$ and $\beta'$ are given in Table \ref{Tab:PropConst2}. For  $\lambda_0=852$ nm this fiber can support four guided modes and comparing the rates obtained for $l=200$ nm  with the ones for the single-mode fiber (solid line in Figs.~\ref{fig:changingd} and ~\ref{fig:changinga}), one can see that the collection rate increases from about 0.06 to 0.07. This increase persists at the larger distance of $l=250$ nm (dashed lines), however less dramatically and by $l=300$ nm (starred line) it is almost impossible to distinguish the atoms. While using a larger fiber gives an advantage in terms of radiation collected,  Fig.~\ref{fig:fourmodes} also shows that the visibility goes down faster and therefore the resolution for single lattice sites decreases.

\begin{figure}[t]
\centering
\includegraphics[width=0.9\linewidth]{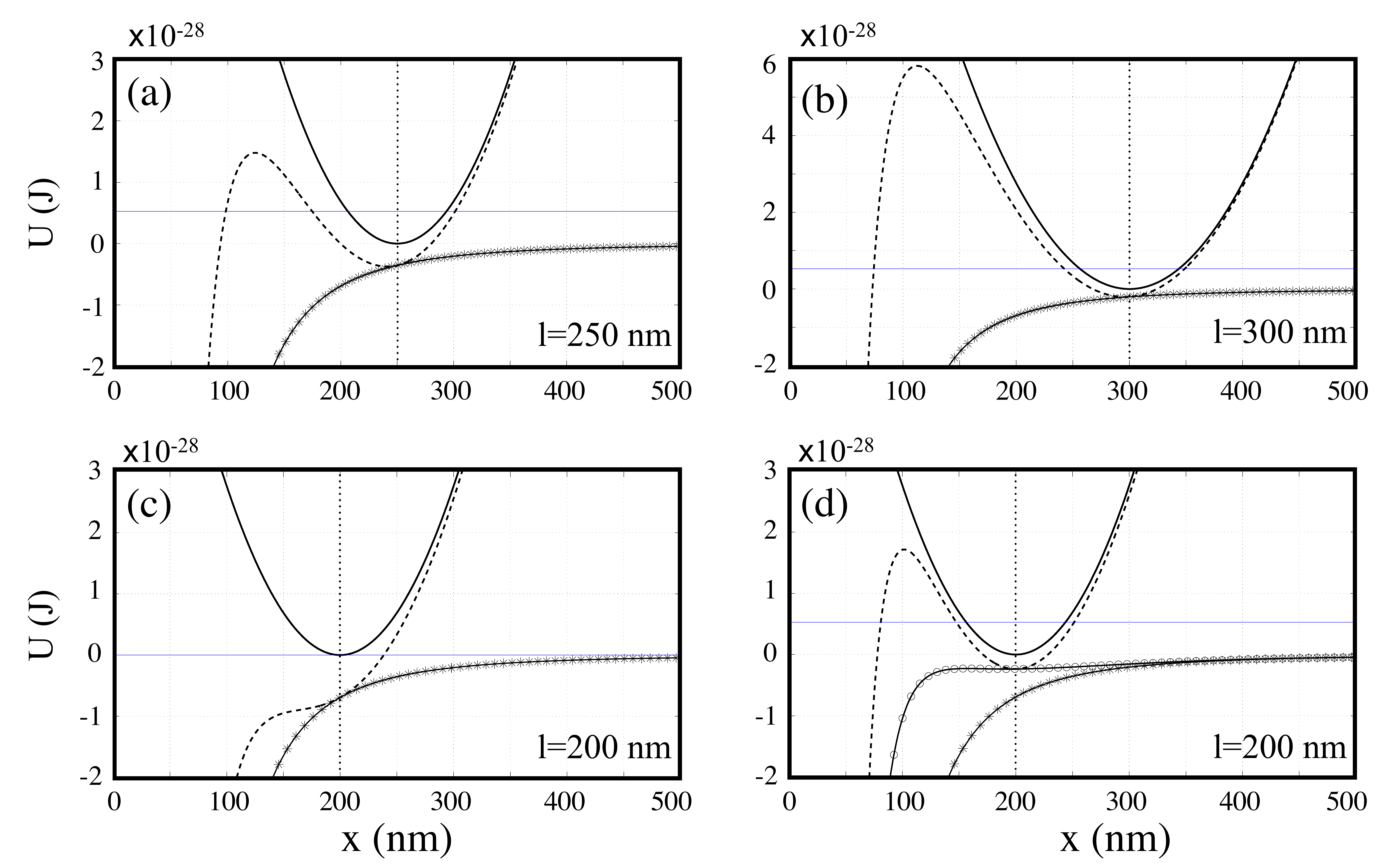} 
\caption{Potentials along the shortest distance between the fiber surface and the atom. The solid line represents the undisturbed harmonic oscillator potential assumed to have a frequency of $\omega$=500 kHz and the starred line is the van der Waals potential. The dashed line is the combined potential of all individual ones. (a) and (b) show that for $l\ge 250$~nm the trapping site is stable, whereas it can be seen in (c) that for $l=200$~nm the minimum of the joint potential is lost. In (d) a blue-detuned field has been added to the fiber to compensate the attractive van der Waals force and the circled line shows the combined van der Waals and blue-detuned potential. One can see that this allows for the restoration of the trapping site.}  
\label{fig:vdwdest}
\end{figure}

From the above it is clear that close proximity between the fiber and the atoms ($< 300$ nm) is advantageous to obtain clearly distinguishable signals. However, bringing a room-temperature fiber close to a trapped ultracold atom requires the consideration of the van der Waals interaction between the atom and the fiber surface. This is an attractive interaction that can influence the position of the trapping minimum and therefore the distance between the atom and the fiber. While this can lead to a different  effective the emission rate into the fiber, the more dramatic effect is that the atom can be lost when the attractive potential destabilises the trapping minimum. The classical van der Waals potential felt by an atom near the surface of a dielectric fiber of infinite length was calculated by Boustimi {\it et al.} \cite{Boustimi:02}, and a detailed analysis of their expression by Le Kien {\it et al.} \cite{LeKien:04} showed that for atoms close to the surface the van der Waals potential tends to the same values as that for a flat surface
\begin{equation}
  \label{eq:vdWflat}
  V_\text{flat}
       =-\frac{1}{(r-a)^3}\frac{\hbar}{16 \pi^2 \epsilon_0}\int_0^\infty \! d\xi\; \alpha(i \xi)
       =-\frac{C_3}{(r-a)^3}\;.    
\end{equation}
For atoms further away, the expression for a flat surface offers an upper limit on the influence of the van der Waals interaction and taking $\lambda_{0}=852$ nm gives a van der Waals constant for $^{133}$CS of $C_3 \approx 5.6 \times 10^{-49}$  J m$^3$ \cite{LeKien:04}.
The combined potentials the atoms sees at two different distances from the fiber ($l=250$~nm and $l=300$~nm) are shown
in Figs.~\ref{fig:vdwdest}(a) and (b), where we have assumed that the atom is transversally trapped in a tight harmonic oscillator trap of frequency of 500 kHz. One can see that at these distances the van der Waals interaction does not significantly effect the trapping position and no corrections to the emission rates are necessary. This, however, is different when the atom comes closer to the surface and for distances around 200 nm, the trapping potential becomes unstable and the atom is lost to the strong attractive potential from the fiber (see Fig.~\ref{fig:vdwdest}(c)). To avoid this and to stabilise the trapping potential, one can add another repulsive field to the fiber, which is blue-detuned for the atom and its evanescent field partly compensates the van der Waals potential. This is is demonstrated in Fig.~\ref{fig:vdwdest}(d), where a blue-detuned field of wavelength $\lambda_b=440$ nm and intensity $P_b=1.75$ mW is added to the fibre and allows to restore the original trapping location.

However, this compensation mechanism does not work for arbitrarily short distances, due to the different functional forms of the van der Waals and the optical force. And while we have seen that shorter distances give better resolution, the van der Waals attraction sets an upper limit to what is achievable.

\section{Limits} 

To demonstrate the limits of the technique proposed above and to apply it to a relevant dynamical situation, we return to considering a single-mode fiber of radius $150$ nm in this section. From the results shown in Fig.~\ref{fig:changingd} we have seen that a nanofiber with a radius smaller than half an optical wavelength can resolve the position space of a collection of atoms trapped in an optical lattice. It is therefore a natural question to ask what the limit of this approach is and if it can be used to resolve more complicated atomic distributions or dynamical processes.
 
\begin{figure}[H]
  \centering
  \begin{subfigure}{.38\textwidth}
    \centering
\includegraphics[width=0.95\linewidth ]{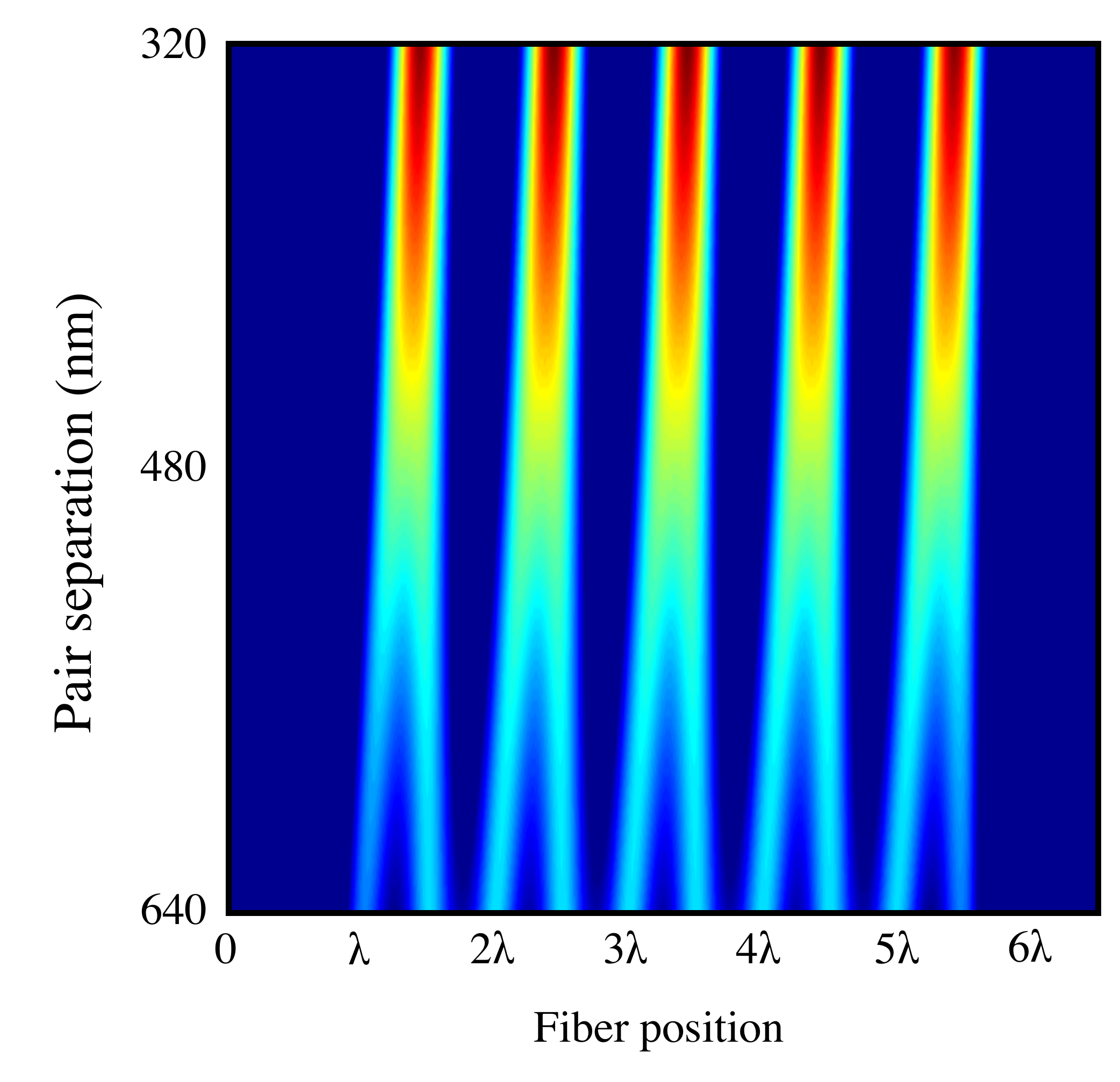} 
    \caption{}
\label{fig:pcolorasatomspair}
  \end{subfigure}%
  \begin{subfigure}{0.62 \linewidth}
    \centering
\includegraphics[width=\linewidth]{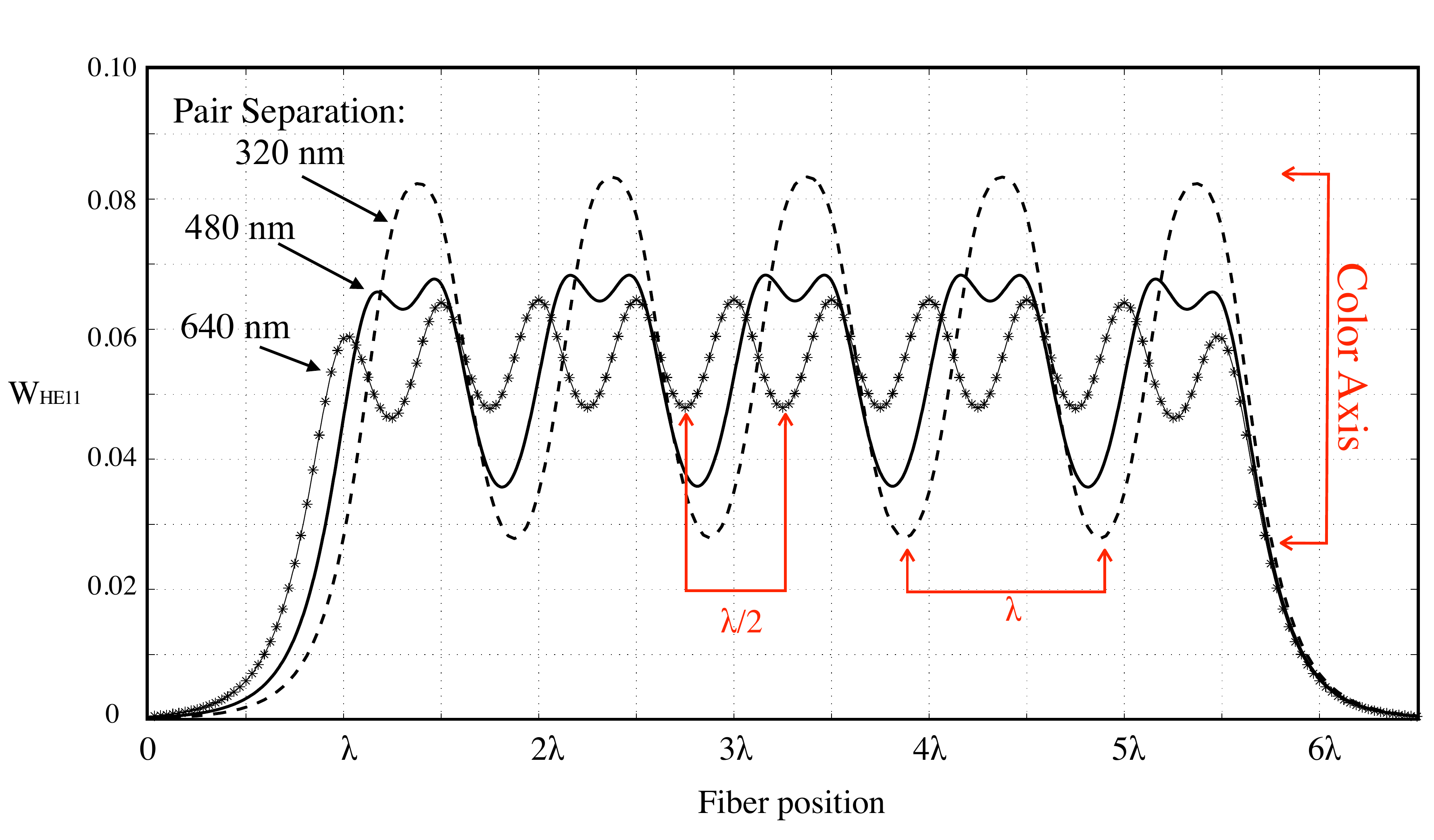} 
        \caption{}
\label{fig:pcolorasatomspair2}
  \end{subfigure}
 \caption{ (a) At the bottom of this figure, 10 atoms are arranged in a row, equidistant from one another with a spacing of $\lambda/2=640$~nm. Moving towards the top, every other atom shifts closer to its neighbour on the right. The emission from two different atoms can be distinguished until the their separation is closer than approximately $ 400$~nm. (b) Here we show three slices from Fig.~\ref{fig:pcolorasatomspair} when the pair separation is equal to 640 nm (starred line),  480 nm (solid line) and  320 nm (dashed line). We also indicate the colorbar axis in Fig.~\ref{fig:pcolorasatomspair}.}
\end{figure}

With this in mind, we show in Figs.~\ref{fig:pcolorasatomspair} and \ref{fig:pcolorasatomspair2} the results for a situation typical in the process of controlled collisions between neutral atoms \cite{Jaksch:1999}, in which every second atom in a row is brought closer to its neighbour. We consider a row of ten atoms initially equally spaced and calculate the spatially resolved emission rate as the distance between pairs becomes smaller using a fiber of radius $a=150$~nm, which is $l=200$~nm away from the row of atoms. One can clearly see that initially the expected well resolved maxima appear, but with decreasing separation between each pair of atoms the respective maxima move closer together and eventually the pairs of atoms become indistinguishable from each other. A simple geometrical argument based on the dipole emission cones shows that this should happen at when the separation between the pairs is approximately $2l$, for a typical fiber radius, which is confirmed by the graph. However, as the intensity measured in the fiber increases, one can still infer the presence of two atoms.

Let us finally remark that even though our calculations have been done for a one-dimensional setting, one can easily imagine using a fiber to measure states at the edge of a two-dimensional geometry.  Moving the nanofiber through a two-dimensional lattice, however, would lead to losses due to the finite reach of the van der Waals potential and in \cite{Hennessy:12} it was shown that compensation via a blue-detuned field can compensate well enough to allow the fiber to be moved through a lattice while compromising only the very closest sites to itself. This still allows, for example, to make measurements on alternate rows. While the analysis of the emission spectrum would be more difficult as emitters are located in two-dimensional space, one can compare measurements from different positions in order to determine the occupation of single lattice sites or even employ fiber arrays \cite{NicChormaicPrivateCommunication}.

\section{Conclusions}

We have shown that subwavelength optical nanofibers can be used to resolve atomic distributions in optical lattices. This is an alternative approach to the recently developed atom microscope and allows for higher resolution when using smaller fibers. However, as it requires bringing macroscopic objects very close to single atoms, it is mainly limited to one-dimensional arrays or edges of higher dimensional ones. Nevertheless, it offers  a new approach that can be integrated in ultracold atom experiments with current technologies.

\section{Acknowledgements}
We would like to thank C.F.~Phelan for valuable discussions. This work is dedicated to the memory of Prof.~V.~Minogin.

\section*{Appendix}

The explicit forms for the emission rates into the individual modes of the fiber given in Sec.~\ref{sec:EmissionRates} contain the following expressions to be completed

\subsection*{HE$_{11}$ MODE}

\begin{align}
N_1= & J_1^2(h a) - {J_0} (h a){J_2} (h a)+\frac{\beta^2}{2 h^2}\Big[ (1-s)^2 \left(J_0^2(ha)+J_1^2(ha)\right)\nonumber\\
&\qquad\qquad\qquad\qquad\qquad\qquad\qquad+(1+s)^2 \left(J_2^2(ha)+J_1(ha) J_3(ha) \right) \Big]\\
N_2=&\frac{J_1^2 (h a)}{K_1^2 (q a)}\Bigg[ K_0 (q a)K_2 (q a)- K_1^2(q a) 
+\frac{\beta^2}{2 q^2}\Big[ (1-s)^2 \left(K_1^2(qa)+K_0^2(qa)\right)\nonumber \\
&\qquad\qquad\qquad\qquad\qquad\qquad\qquad+(1+s)^2 \left(K_1(qa) K_3(qa) -K_2^2(qa)\right) \Big]\Bigg]\\
s=&\frac{{1/{h^2a^2}}+{1/{q^2a^2}}}{{{{J_1'(h a)}/ {{h a J_1(h a)}}}+{{K_1'(q a)}/{q a K_1(q a)}}}}
\end{align}
\\
\subsection*{TE$_{01}$ MODE}

\begin{align}
P_1=&\frac{1}{a^2h^2}\frac{K_0^2 (q a)}{J_0^2 (h a)}\left(J_1^2 (h a)- J_0 (h a)J_2 (h a)\right)\\
P_2=&\frac{1}{a^2q^2}\left(K_0 (q a)K_2 (q a)-  K_1^2 (q a)\right)
\end{align}
\\
\subsection*{TM$_{01}$ MODE}

\begin{align}
Q_1=&\frac{K_0^2 (q a)}{J_0^2 (h a)} \left[ J_0^2(ha) + \frac{n_1^2 k^2}{h^2} J_1^2(ha) -\frac{\beta^2}{h^2} J_0 (h a)J_2 (h a)  \right]\\
Q_2=&\frac{\beta^2}{q^2}   K_0 (q a)K_2 (q a)  -K_0^2 (q a) - \frac{n_2^2 k^2}{q^2} K_1^2 (q a) 
\end{align}
\\
\subsection*{HE$_{21}$ MODE}

\begin{align}
R_1=& J_2^2 (h a) -{J_1} (h a){J_3} (h a) +\frac{\beta^2}{2 h^2} \Big[ (1-u)^2( J_1^2 (h a)- {J_0} (h a){J_2} (h a))  \nonumber\\ 
&\qquad\qquad\quad\qquad\qquad\qquad\qquad+  (1+u)^2( J_3^2 (h a)- {J_2} (h a){J_4} (h a))\Big]\\
R_2=&\frac{J_2^2(ha)}{K_2^2(qa)}\Bigg[K_1(qa)K_3(qa)-K_2^2(qa)+\frac{\beta^2}{2q^2}\Big[(1-u)^2(K_0(qa)K_2(qa)-K_1^2 (qa))\nonumber\\ 
&\qquad\qquad\quad\qquad\qquad\qquad\qquad+(1+u)^2(K_2(qa)K_4(qa)-K_3^2 (qa) ) \Big] \Bigg]\\
u=&\frac{2\left( {{1/{h^2a^2}}+{1/{q^2a^2}}}\right)} {{{{J_2'(h a)}/ {{h a J_2(h a)}}}+{{K_2'(q a)}/{q a K_2(q a)}}}}
\end{align}


\begin{thebibliography}{99}




\bibitem{Meschede:09} M.~Karski, L.~F\"orster, J.~M.~Choi, W.~Alt, A.~Widera, and D.~Meschede,``Nearest-Neighbor Detection of Atoms in a 1D Optical Lattice by Fluorescence Imaging,''  Phys. Rev. Lett. \textbf{102}, 053001 (2009).

\bibitem{Bakr:09} W.~S.~Bakr, J.~I.~Gillen, A.~Peng, S.~Fölling and M.~Greiner,``A quantum gas microscope for detecting single atoms in a Hubbard-regime optical lattice,'' Nature \textbf{462}, 74-77 (2009).

\bibitem{ward:2014} J.M.~Ward,~Vu~H.~Le, A.~Maimaiti and S.~Nic.~Chormaic, ``Optical micro- and nanofiber pulling rig,'' arxiv.org/abs/1402.6396 (2014)

\bibitem{two atoms} Fam~Le~Kien, S.~Dutta~Gupta, K.~P.~Nayak and K.~Hakuta, ``Nanofiber-mediated radiative transfer between two distant atoms,'' Phys. Rev. A \textbf{72}, 063815 (2005).

\bibitem{NayakHakuta:08} K.~P.~Nayak and K.~Hakuta,``Single atoms on an optical nanofiber,''  New Journal of Physics, \textbf{10} (2008) 053003 (9pp).

\bibitem{6Hakuta:07} K.~P.~Nayak, P.~N.~Melentiev, M.~Morinaga, Fam~Le~Kien, V.~I.~Balykin and K.~Hakuta, ``Optical nanofiber as an efficient tool for manipulating and probing atomic fluorescence,''  Optics Express, Vol. \textbf{15}, Issue 9, pp. 5431-5438 (2007).

\bibitem{Yalla:12} Ramachanndrarrao Yalla, Fam Le Kien, M. Morinaga, and K. Hakuta, ``Efficient Channeling of Fluoresence Photons from Single Quantum Dots into Guided Modes of Optical nanofiber,'' Phys.~Rev.~Lett.~ {\bf 109}, 063602 (2012).

\bibitem{Fujiwara:11} Masazumi~Fujiwara, Kiyota~Toubaru, Tetsuya~Noda, Hong-Quan~Zhao, and Shigeki~Takeuchi, ``Highly Efficient Coupling of Photons from Nanoemitters into Single-Mode Optical Fibers,''  Nano Lett. 2011, 11 (10), pp 4362–4365.

\bibitem{multiatom} F.~Le~Kien and K.~Hakuta, `Cooperative enhancement of channeling of emission from atoms into a nanofiber,'' Adv. Nat. Sci.: Nanosci. Nanotechnol.  \textbf{3} (2012) 035001

\bibitem{sondergaard:01} T.~S\o ndergaard and B.~Tromborg, ``General theory for spontaneous emission in active dielectric microstructures: Example of a fiber amplifier,''  Phys.~Rev.~A {\bf 64}, 033812 (2001).

\bibitem{KienHakuta:05} Fam~Le~Kien, S.~Dutta~Gupta, V.~I.~Balykin and K.~Hakuta, ``Spontaneous emission of a cesium atom near a nanofiber: Efficient coupling of light to guided modes,'' Phys.~Rev.~A, \textbf{72}, 032509 (2005).

\bibitem{Masalov:13} A.V.~Masalov and V.G.~Minogin, ``Pumping of higher-modes of an optical nanofiber by laser excited atoms," Laser Phys.~Lett.~{\bf 10}, 075203 (2013). 

\bibitem{nicchor:14-2} R.~Kumar, V.~Gokhroo, A.~Maimaiti, K.~Deasy, M.~C.~Frawley, S.~ Nic~Chormaic, ``Interaction of laser-cooled 87Rb atoms with higher order modes of an optical nanofiber,''  arXiv:1311.6860

\bibitem{topedgestates} M.Z. Hasan and C.L. Kane,``Colloquium: Topological insulators,''  Rev. Mod. Phys. {\bf{82}} 3045, (2010).

\bibitem{latticebook} \textit{Ultracold atoms in optical lattices:simulating quantum many-body systems.} M.~Lewenstein,~A.~Sanpera,~V.~Ahufinger. Oxford University Press (2012).

\bibitem{Jackson} See, for example, J.D. Jackson, \textit{Classical  Electrodynamics}, 3rd ed. (John Wiley \& Sons, New York, 1998).

\bibitem{Greiner:02} M.~Greiner, O.~Mandel, T.~Esslinger, T.W.~H\"ansch, and I.~Bloch, ``Quantum phase transition from a superfluid to a Mott insulator in a gas of ultracold atoms,'' Nature {\bf 415} 39, (2002).

\bibitem{Becker:09} C.~Becker, P.~Soltan-Panahi, J.~Kronj\"ager, S.~D\"orscher, K.~Bongs, and K.~Sengstock, ``Ultracold quantum gases in triangular optical lattices,'' arXiv:0912.3646 (2009).

\bibitem{Tong:03} L.~Tong, R.R.~Gattass, J.B.~Ashcom, S.~He, J.~Lou, M.~Shen, I.~Maxwell, and E.~Mazur, ``Subwavelength-diameter silica wires for low-loss optical wave guiding,'' Nature {\bf 426}, 816 (2003).

\bibitem{Vetsch:09} E.~Vetsch, D.~Reitz, G.~Sagu\'e, R.~Schmidt,S.T.~Dawkins, A.~Rauschenbeutel, `` Optical interface created by laser-cooled atoms trapped in the evanescent field surrounding an optical nanofiber,'' Phys. Rev. Lett. \textbf{104}, 203603 (2010) 

\bibitem{Morrissey:09} M.J.~Morrissey, K.~Deasy, Y.~Wu, S.~Chakrabarti and S.~Nic Chormaic, ``Tapered optical fibers as tools for probing magneto-optical trap characteristics,'' Rev.~Sci.~Instrum.~{\bf 80}, 053102 (2009).




\bibitem{Sellmeier} The refractive index $n_1$ of fused silica
  $(SiO_2)$ can be calculated using a Sellmeier-type dispersion formula
 , taking the refractive index of the vacuum $n_2 =1$
\begin{align}
  n_1-1=&\frac{0.696166\lambda^2}{\lambda^2-(0.068404)^2}
        +\frac{0.407942\lambda^2}{\lambda^2-(0.116241)^2}\nonumber\\
       &+\frac{0.897479\lambda^2}{\lambda^2-(9.896161)^2}\nonumber
\end{align}
where $\lambda$ is in units of $\mu$m.

\bibitem{fiber books} See, for example, 
D. Marcuse, \textit{Light Transmission Optics} 
(Krieger, Malabar, FL, 1989);
A. W. Snyder and J. D. Love, \textit{Optical Waveguide Theory} (Chapman and Hall, New York, 1983).

\bibitem{Boustimi:02} M.~Boustimi, J.~Baudon, P.~Candori, and J.~Robert, ``van der Waals interaction between an atom and a metallic nanowire,'' Phys.~Rev.~B {\bf 65}, 155402 (2002).

\bibitem{LeKien:04}F.~Le Kien, V.I.~Balykin, and K.~Hakuta, ``Atom trap and waveguide using a two-color evanescent light field  around a subwavelength-diameter optical fiber,"  Phys.~Rev.~A {\textbf 70}, 063403 (2004).

\bibitem{Jaksch:1999} D.~Jaksch, H.J.~Briegel,~J.I.~Cirac,~C.W.~Gardiner, and P. Zoller, ``Entanglement of atoms via cold controlled collisions,''  Phys. Rev. Lett. \textbf{82}, 1975–1978 (1999).

\bibitem{Hennessy:12} T.~Hennessy and Th.~Busch, ``Creating atom-number states around tapered optical fibers by loading from an optical lattice,"
Phys.~Rev.~A \textbf{85}, 053418 (2012).

\bibitem{NicChormaicPrivateCommunication} S.~Nic Chormaic, private communication.
\end{thebibliography}
\end{document}